# A High-Responsivity Broadband Photodetector Based on a WSe$_2$/NiO Nanowire Heterostructure with Engineered Nanophotonic Enhancement

Chandra Sekhar Reddy Kolli[1], Gowtham Polumati[1], Aleksandra A. Kutuzova[2], Ekaterina E. Maslova[2], Andres De Luna Bugallo[3], Nitish Kumar Gupta[1], Mikhail V. Rybin[2,4], Parikshit Sahatiya[1*]

[1] *Department of Electrical and Electronics Engineering, BITS Pilani, Hyderabad Campus, Hyderabad, 500078, India.*

[2] *ITMO University, St. Petersburg 197101, Russia*

[3] *Centro de Física Aplicada y Tecnología Avanzada, Universidad Nacional Autónoma de México, A.P. 1-1010, Querétaro, Qro., C.P. 76000, México*

[4] *Ioffe Institute, St. Petersburg 194021, Russia*

*Corresponding author E-mail: parikshit@hyderabad.bits-pilani.ac.in.

**Abstract:**

Engineering nanoscale light–matter interaction in mixed-dimensional semiconductor heterostructures provides a pathway to mitigate the intrinsic gain–bandwidth trade-off in photodetectors. Here, we report a broadband, high-responsivity 2D/1D photodetector formed by integrating monolayer p-type WSe$_2$ with electrospun p-type NiO nanowires. We demonstrate that the device photoresponse, spanning the ultraviolet to near-infrared range (350–780 nm), is governed by a nanophotonic field-confinement mechanism rather than by bulk optical absorption alone. The high-index NiO nanowire functions as a dielectric Mie-type nanoresonator that supports geometry-defined optical modes and produces antenna-like near-field concentration of the optical electric field at the nanoscale WSe$_2$/NiO junction. This localized optical mode enhances the local absorption cross-section and thus the photocarrier generation rate within the junction region, which we identify as the dominant active volume for photocurrent formation. A coupled optoelectronic model, linking full-wave solutions of Maxwell's equations (spatially integrated near-field energy density, $\int|E|^2 dV$) to carrier generation, recombination, and extraction, quantitatively captures both the measured responsivity spectrum and its incident-power dependence using only two electronic fitting parameters. The device achieves responsivities of 627 A/W (visible), 227 A/W (UV), and 167

A/W (NIR), demonstrating simultaneous broadband operation and ultrahigh gain. These results indicate that mixed-dimensional 2D/1D semiconductor junctions can operate as self-aligned nanoscale optical antenna–absorber systems, in which photonic near-field concentration at the junction directly drives high carrier generation rates and thus large external gain. This establishes geometric resonance in mixed-dimensional systems as a generalizable and powerful design principle for developing next-generation, high-gain optoelectronic detectors.

## 1. Introduction

The rapid expansion of applications, such as high-speed optical communication, biomedical imaging, autonomous sensing, and quantum technologies, has created a pressing demand for photodetectors that can simultaneously deliver high responsivity and broad spectral coverage [1-3]. Yet, conventional semiconductor-based devices remain fundamentally constrained by a trade-off: achieving high sensitivity often restricts their operational bandwidth, while extending the spectral range typically reduces efficiency. This bottleneck limits their practical applicability in next-generation optoelectronic systems, where multifunctional performance is critical [4-5]. Overcoming this challenge requires new material platforms and design strategies that can manipulate light–matter interactions beyond the limits of bulk semiconductors. In this context, low-dimensional and mixed-dimensional heterostructures have emerged as particularly promising candidates, as their unique geometries and optical responses enable synergistic enhancement of both absorption and carrier dynamics [6-7]. While van der Waals (vdW) heterostructures have been extensively investigated as a pathway to overcome these limitations by exploiting tailored band alignments and interfacial charge transfer, studies on mixed-dimensional systems, particularly 2D/1D heterostructures, remain relatively scarce.

Their limited exploration arises largely from fabrication complexities and challenges in achieving controlled nanoscale interfaces [8-9]. Hence, there is an urgent need to investigate such architectures, as they offer unique opportunities to synergistically couple electronic and photonic effects for advancing photodetector performance

Two-dimensional (2D) semiconductors have emerged as a versatile platform for optoelectronics due to their atomically thin geometry, high surface-to-volume ratio, and strong excitonic effects, which collectively enable efficient light–matter interactions across a wide spectral range[10-12]. When integrated into vdW heterostructures, they offer additional tunability by coupling with materials of different dimensionalities without the constraints of lattice matching[13-15]. Depending on the partner dimensionality, distinct photonic properties arise, which are 2D/2D heterostructures benefit from ultraclean band alignment and efficient charge transfer but suffer from limited optical path length and weak field confinement due to their planar nature; 2D/0D hybrids leverage quantum confinement and discrete energy states of quantum dots to extend absorption spectra, though interfacial trap states often hinder efficient exciton dissociation; and 2D/3D systems combine strong bulk absorption with atomically thin channels, but the large mismatch in dimensionalities can dilute local field enhancement at the junction[16-17]. In contrast, 2D/1D mixed-dimensional heterostructures uniquely integrate the exciton-rich response of 2D semiconductors with the resonant photonic functionality of 1D nanostructures[18]. The subwavelength dimensions of nanowires or nanotubes enable them to act as dielectric antennas supporting Mie-type modes, producing strong electric field confinement and significantly enhancing the absorption cross-section at the junction[19-21]. This synergistic interplay between electronic and photonic effects makes 2D/1D heterostructures particularly powerful for overcoming the long-standing trade-off between responsivity and spectral bandwidth in photodetectors[22-24].

Here, we report on a mixed-dimensional broadband photodetector realized by integrating 1D p-type NiO nanowires with 2D p-type WSe$_2$ monolayers grown by chemical vapor deposition on the Si/SiO$_2$ substrate. The device is fabricated by precisely placing electrospun single NiO nanowires across pre-patterned contacts on the WSe$_2$ channel, forming a 1D/2D heterojunction that enables efficient charge carrier collection at the respective terminals under illumination. The heterostructure exhibits consistently high responsivities of 627 A/W, 227 A/W, and 167 A/W under visible, ultraviolet, and near-infrared excitations, respectively, across a broad spectral range of 350–780 nm. The exceptional broadband response originates from a nanophotonic enhancement mechanism, wherein the subwavelength NiO nanowire acts as a resonant dielectric antenna supporting Mie-type modes. This resonant field localization amplifies the absorption cross-section at the 1D/2D interface, thereby enhancing photocarrier generation and establishing geometric resonance as the dominant operative pathway. The coupled optoelectronic modeling further corroborates the experimental findings, quantitatively reproducing the measured spectral responsivity.

## 2. Experimental section:

***Synthesis of 2D-WSe$_2$ flakes:*** Initially, tungsten trioxide (WO$_3$) and selenium (Se) powder of 20 mg and 25 mg, respectively, were kept in two alumina oxide (Al$_2$O$_3$) boats as shown in **Fig.1a**. A Si/SiO$_2$ wafer was cleaned using an RCA cleaning process, and then the Si/SiO$_2$ wafer with dimension 1 cm × 2 cm was placed Si face down on the Al$_2$O$_3$ boat with WO$_3$. Before the deposition, a vacuum was created in a quartz tube using a mechanical pump. Later, the quartz tube was purged with Ar + H$_2$ gas, and it was used as carrier gas at a flow rate of 40 sccm. During the reaction, the temperature of the WO$_3$ and Se zones was maintained at 850°C and 750°C, respectively, and both WO$_3$ and Se zones were heated simultaneously. The growth temperature was maintained for 10 minutes, and the quartz tube was subsequently cooled down to room temperature. Metallic alignment markings were employed on top of the WSe$_2$ crystals.

The contact pattern was then carried out by an electron-beam lithography. Cr (10 nm)/Au (200 nm) metallization was coated by thermal evaporation followed by a lift-off method to finalize the device fabrication, as shown in **Fig 1e**.

***Synthesis of 1D-NiO NW***: Polyvinyl alcohol (PVA, MW = 125,000) at a concentration of 10 wt% was combined with 10 ml of deionized water and agitated for 30 minutes. Additionally, 0.1 M of a 10 wt% $Ni(NO_3)_2 \cdot 6H_2O$ precursor was included directly into the polymer solution. The solution was agitated for 12 hours to get a thick green consistency. The precursor solutions were subsequently transferred to the electrospinning syringe, which had an inner needle diameter of 0.34 mm. The solution flow rate was 0.6 ml/h, and the electrospun fibers were originally collected and verified by positioning an ITO/PET substrate (trial substrate) adhered to aluminum foil, located 15 cm from the syringe tip. A DC voltage of 12 kV was given to the metal needle to generate a Taylor cone fibrous structure from the tip [25]. To construct the 1D p-NiO/2D-p-WSe2 heterojunction, the trial substrate was substituted with a CVD-grown 2D-WSe2 device (with contacts prepared using electron beam lithography) that was affixed to the electrospinning collector to gather a single 1D-NiO fiber during an electrospinning duration of one second, as illustrated in **Fig 1d**. The as-deposited fibers were calcined at 450 °C for one hour in air using a standard oven. The highly crystalline 1D p-NiO nanowires were obtained following thermal calcination at 450 °C.

The morphology of both 1D p-NiO nanowires and 2D p-WSe2 was examined using field emission scanning electron microscopy (ZEISS Ultra-55 SEM). The crystallinity of 1D p-NiO was assessed using X-ray diffraction (Rigaku Dmax2100). Thermogravimetric analysis (TGA) was conducted using a Perkin Elmer Pyris 1, employing alumina sample holders, with a heating rate of 10 °C min−1, from ambient temperature to 700 °C under a nitrogen flow. X-ray photoelectron spectroscopy (XPS) investigations of one-dimensional p-NiO nanowires and two-dimensional p-WSe2 elucidated their chemical functional groups and binding energies.

Photodetection experiments and electrical measurements were conducted with a vertical monochromatic light source with a wavelength range of 365 nm to 780 nm, an illumination area of 20 mm², and a four-probe station with a Keithley 2450 Source Meter. A Lab RAMHR Raman spectrometer was utilized to conduct Raman spectral observations.

## 3. Results and Discussion

### 3.1. Material characterization

An optical microscope was used to examine the surface morphology of $WSe_2$ deposited on the $Si/SiO_2$ wafer. It was observed that $WSe_2$ flakes have a triangular shape with good uniformity along with some arbitrary flakes. The size of different flakes was varying between 15-45μm as shown in **Fig.1a.** The examination by Raman spectroscopy revealed a pair of intensity peaks at 250 $cm^{-1}$ and 260 $cm^{-1}$ corresponds to the $A_{1g}$ and $E_{2g}^1$ modes, which are out-of-plane and in-plane vibrations of the Se atoms with respect to W atoms. The frequency difference between in-plan and out-of-plane modes is around 10 $cm^{-1}$, conforming to monolayer $WSe_2$[26]. Moreover, the absence of peaks at 310 $cm^{-1}$ appearing in bilayer or multi-layered flakes further confirms that the flake is a monolayer (**Fig.1b).** The photoluminescence intensity graph shows the peak at ~784 $cm^{-1}$[27], also indicating the monolayer nature of the wafer as shown in **Fig. 1c.** NiO NWs were directly synthesized via electrospinning onto the $WSe_2/Si/SiO_2$ substrate, followed by calcination at 450°C for final device fabrication. Prior to this process, the crystallinity and morphological characteristics of the NiO NWs were analyzed. The FESEM image (**SI, Fig. S1a**) depicts electrospun NWs with diameters of approximately 200 nm and lengths of several micrometers. Additionally, the TEM image confirms their uniform orientation (**SI, inset Fig. S1a**). The strong adherence of NWs to the substrate is attributed to the electrostatic repulsion generated between the collector and the syringe needle during the electrospinning process. Following the calcination process, the NWs remained firmly adhered to the substrate without the need for additional binders or transfer processes [28]. This strong adherence is achieved through two key factors: (i) the controlled viscosity of the polymer-Ni precursor mixture, and (ii) the precise tuning of deposition time and rate during the electrospinning process.

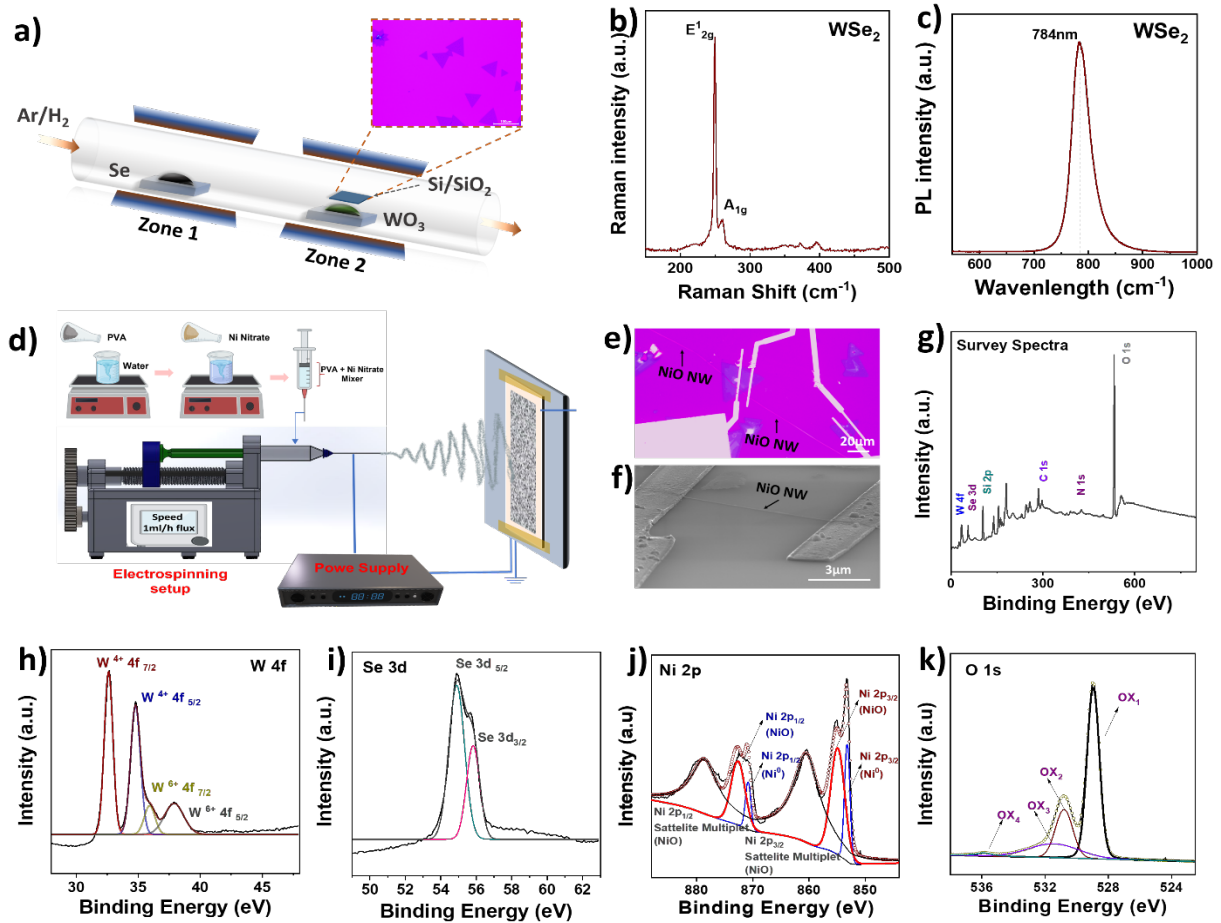

**Fig:1** Schematic showing a) complete synthesis of CVD-grown monolayer $WSe_2$; b) Raman spectrum of $WSe_2$; c) PL spectrum of $WSe_2$; d) Electrospinning of 1D p-NiO; e-f) Optical image and FESEM image of e-beam contacts between p-$WSe_2$ flake and p-NiO nanowire; g) XPS survey spectra of $WSe_2$ deposited on Si wafer; h-i) Narrowband spectrum of W 4f and Se 3d. j-k) XPS analysis with Gaussian deconvoluted Ni 2p and O1s spectra.

Furthermore, the chemical composition of as-grown WSe2 flake is examined using X-ray photoelectron spectroscopy (XPS) analysis, and the spectra confirm the presence of W, Se, and oxygen (due to $SiO_2$ on Si) elements as shown in **Fig.1 g).** High resolution spectra of both W

4f and Se 3d were included where in W 4f shows peaks located at 32.5 eV (W $4f^{7/2}$) and 35.0 eV (W $4f^{5/2}$) respectively (**Fig.1h**). Similarly, the Se peaks are located at 54.9 eV (Se $3d^{5/2}$) and 55.7 eV (Se $3d^{3/2}$) respectively (**Fig.1i**). These results are consistent with previously reported values[29-30]. In addition, we observed a weak peak around 35 eV in the XPS spectra of W 4f. This could be attributed to an oxidation reaction during sample preparation as $WSe_2$ was grown by selenization of $WO_3$. Further, the purity of NiO nanowires and the electronic bonding states of the calcined nanowires were studied using XPS analysis, with Gaussian-deconvoluted Ni 2p and O 1s spectra presented in **Fig.1j-k.** The Ni 2p spectrum depicted in **Fig. 1j** reveals two prominent spin-orbit multiplet peaks: Ni $2p^{3/2}$ at 853.8 eV and Ni $2p^{1/2}$ at 855.7 eV, along with their respective satellite peaks at 881.5 eV and 860.8 eV. The Ni $2p^{1/2}$ multiplet at 853.3 eV indicates a typical Ni-O octahedral bonding arrangement in cubic phase NiO. The Ni $2p^{3/2}$ is congruent with the documented NiO spectrum. Additionally, a peak at 872.4 eV is associated with the $Ni^{2+}$ vacancy caused by the $Ni^{3+}$ ion or $Ni(OH)_2$. The O1s XPS spectrum (depicted in **Fig. 1k**) was deconvoluted into four peaks, with the most prominent peak at 529.1 eV corresponding to Ni-O octahedral bonding. The peaks at 531.6 eV and 532.9 eV correspond to $Ni(OH)_2$ and oxyhydroxide groups, respectively. These results align with [25], corroborated by XRD analysis, and indicate the adsorption of hydroxyl groups on the nanowire surface, as well as the emergence of nickel vacancies within the NiO lattice during heat treatment. No more contaminants were identified, demonstrating the formation of pure NiO nanowires. The electrospun nanofibers are further characterised by XRD as shown in **Fig. S1a of SI** (supplementary Information), wherein it clearly reveals the amorphous structure due to the presence of a mixed polymer composite at room temperature. The XRD pattern of the PVA/Ni $(NO_3)_2$ nanofibers exhibits diffraction peaks at 37.2°, 43.2°, and 62.8°, corresponding to the (1 1 1), (2 0 0), and (2 2 0) crystallographic planes of the cubic NiO phase [31]. The calcination process at 450 °C guarantees the total removal of the polymer from the nanofibers, as corroborated by the TGA curve presented in **Fig. S1b of SI.** The absorbance spectra of p-NiO nano fibre shown in **Fig S1c of SI** clearly indicate that the maximum absorbance corresponds to the UV range[32].

### 3.2. Electrical characterization:

A photodetection experiment was conducted to examine the light matter interaction by subjecting the device to UV, Visible, and NIR light illumination with wavelengths ranging from 365 nm to 780 nm. Prior to the experiment, extreme care was taken to properly place the device in a dark environment to avoid light interaction for accurate experimental results. **Fig.2a** shows the device schematic illustrating a monolayer 2D p-WSe2 flake deposited on Si/SiO2 and on top as an electrospun 1D p-NiO nanowire with gold contacts in between. The detailed FESEM images show e-beam contacts between p-WSe$_2$ flake and p-NiO nanowire, as shown in Fig. **S2a-f** of the **SI**. The device's electrical behavior (I-V sweeps) was obtained by sweeping the applied voltage under each illumination condition. Temporal response traces were recorded at a fixed operating bias of 2 V while varying the optical intensity [**Fig 2b-d)].** The device temporal response was observed by subjecting to UV Visible and IR light illumination with constant intensity (black lines,0.17mW/cm$^2$) and increasing intensities (red lines, 0.17mW/cm$^2$, 0.36mW/cm$^2$, 1.56mW/cm$^2$, 6.36mW/cm$^2$) wherein it was observed a proportional increase in device photocurrent in accordance with the increase in intensity when the light source is ON and reaching to its initial current when the light source is OFF as shown in **Fig 2e-g).** The increase in photocurrent with the increase in incident light intensity clearly reveals the efficient carrier generation upon receiving light energy with different light sources and at different intensities.

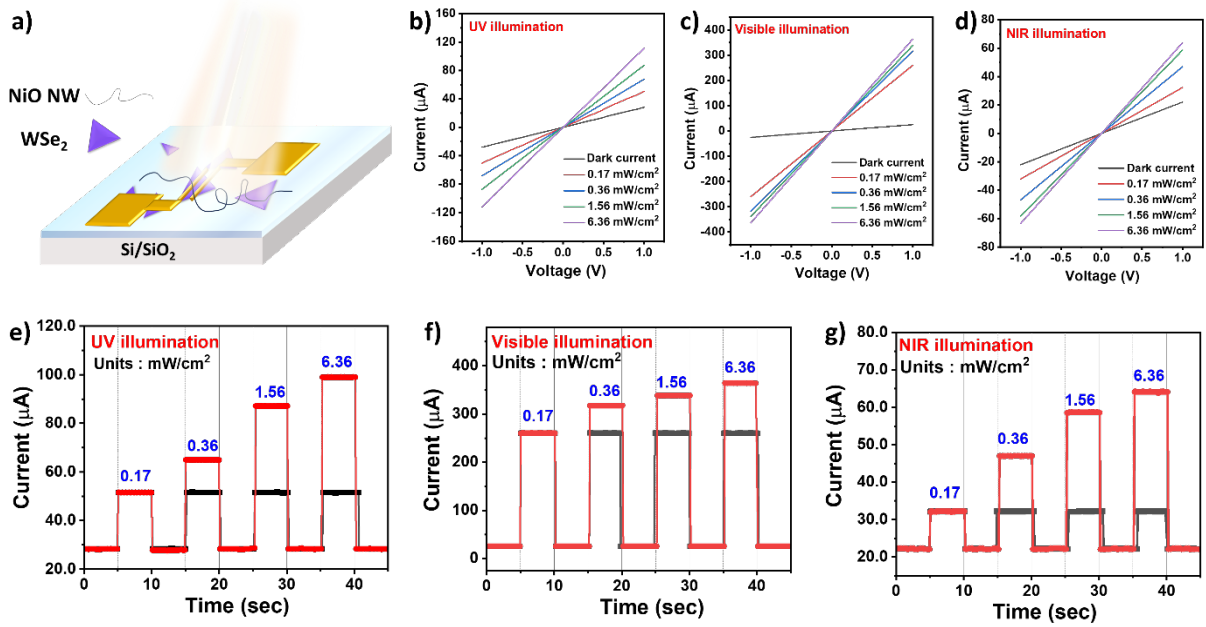

**Fig: 2**a) Schematic of device; b-d) I-V dependences at UV, Visible, and NIR illuminations; e-g) Temporal response with constant intensity and increasing intensity for UV, Visible, and IR light illumination.

Below we deal with three important quantities measuring the photodetector performance, which are responsivity, detectivity and EQE [32]

$$R_\lambda = \frac{I_\lambda}{A*P_\lambda},$$

EQE: hC* ($R_\lambda$/e*$\lambda$) * 100,

$$D^* = (R_\lambda * \sqrt{A})/\sqrt{2*q*I_{dark}},$$

where, $I_\lambda$, $P_\lambda$, $A$, $I_{dark}$, e and $\lambda$ are photo-generated current, incident source power, active area of the device, dark current of the device, charge of electron, wavelength, respectively. Responsivity *R* of a photodetector is a measure of the degree of efficiency with which the incident light energy is converted into its equivalent electrical energy at a given wavelength. Detectivity *D* is defined as the ability of the photodetector device to detect the smallest possible light flux component upon subjecting it to incident light. External Quantum Efficiency (EQE) is the ability of the photodetector device to generate electron-hole pairs for each absorbed photon when subjected to light illumination.

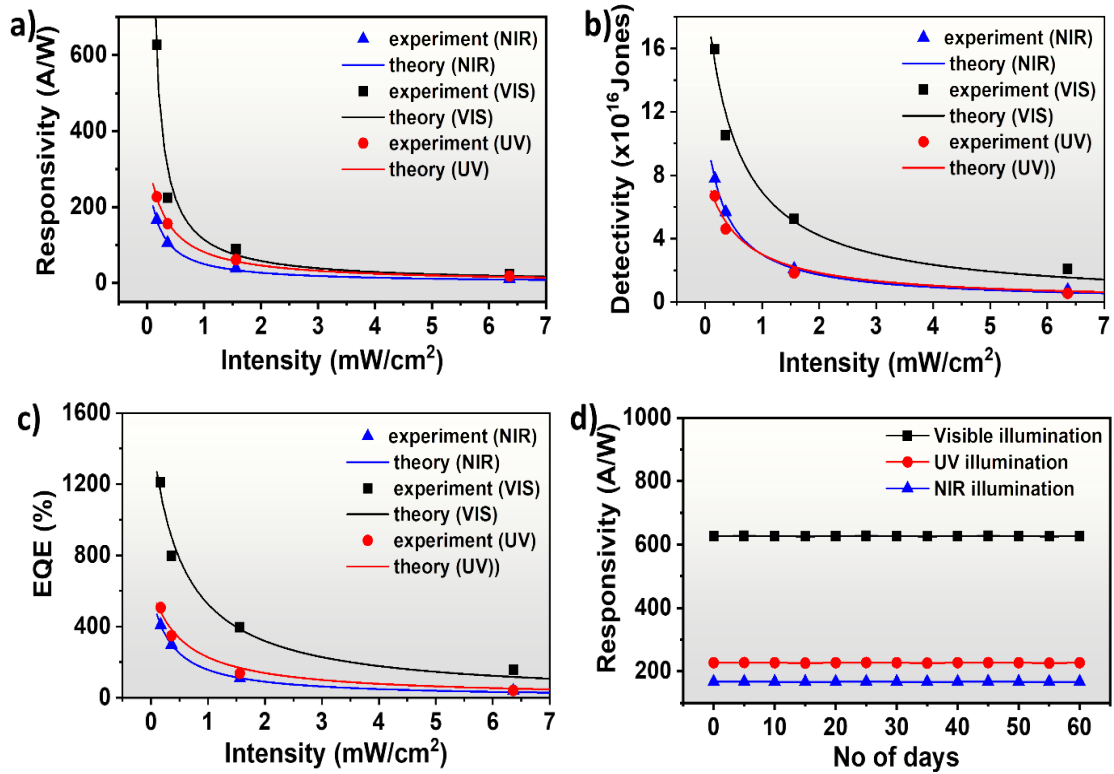

**Fig:3** Graphs showing theoretical and experimental figure of merits of fabricated photodetectors: a) responsivity vs intensity; b) detectivity vs intensity; C) EQE (%) vs intensity; d) stability of the device over 60 days.

Fig. 3a) shows the graph of responsivity as a function of intensity, and the corresponding responsivities are 627 A/W, 227 A/W, and 167 A/W under the low-intensity illumination with Visible, UV, and NIR light incidence, respectively. The high responsivity at Visible incidence is due to the fabricated 2D p-WSe$_2$/1D p-NiO device being occupied with a large, uniform area of moderate band gap 2D p-WSe2 material with high absorption in the visible region and a high concentration of photogenerated carriers. An abrupt decrease in responsivity was also observed when the wavelength was gradually tuned to the NIR region. However, the device has exhibited a moderate responsivity due to moderate absorption of the large band gap 1D p-NiO material shown in Fig. SI 1c). Therefore, from experimental results, it is well understood that the moderate band gap material 2D p-WSe2 shows high absorption at visible and low absorption at NIR, whereas the large band gap material 1D p-NiO shows high absorption at UV. The carrier generations and responsivities exhibited by the 2D p-WSe$_2$/1D p-NiO device are in accordance with their individual absorption capabilities at different illumination sources. Also, a decreasing trend was observed in the responsivity with an increase in light intensity, and the reason was the dominance of the variation of light intensity on the photocurrent. The decrease in responsivity with the increase of incident light intensity was similarly observed with EQE and detectivity. The device has exhibited an EQE (%) of 1403, 507 and 358 and Detectivities of $1.8 \times 10^{17}$, $6.7 \times 10^{16}$ and $4.9 \times 10^{16}$ jones for Visible, UV and NIR illuminations respectively [**Fig 3b-c)**]. The device responsivity was tested for reliability at regular intervals of 10 days for 6 times and noted negligible changes for Visible, UV, and NIR illuminations, respectively, revealing stable performance as shown in **Fig. 3e**).

### 3.3. Photodetection Mechanism:

The operation of the 1D p-NiO/2D p-WSe$_2$ mixed-dimensional heterostructure photodetector can be understood in terms of two complementary mechanisms: (i) carrier generation and transport governed by the band alignment of the constituent semiconductors and the position of the external contacts, and (ii) resonant nanophotonic enhancement arising from the subwavelength geometry of the NiO nanowire. Together, these effects provide a complete picture of the consistently high responsivity observed across UV, Visible, and NIR regions.

**Charge Generation and Transport Dynamics**

**Fig. 4a** schematically illustrates the band diagrams of isolated and contacted 2D p-WSe$_2$ and 1D p-NiO. Ultraviolet photoelectron spectroscopy (UPS) measurements reveal work functions of 4.8 eV for WSe$_2$ [33] and 5.36 eV for NiO [34] (Supplementary Information, **Fig. S3**), which establish a favourable energy offset at the 1D/2D interface. Depending on the spectral region of excitation, photocarrier generation occurs predominantly in one of the two materials, with carrier collection pathways dictated by the relative placement of the contacts.

When the device is illuminated with UV light, the wide bandgap of NiO (~3.4–4.1 eV) enables strong absorption, producing electron–hole pairs within the nanowire[35]. These photogenerated carriers are collected efficiently at the corresponding terminals, as depicted in case (i) of **Fig. 4a**. At visible and NIR wavelengths, absorption is dominated by the WSe$_2$ monolayer due to its moderate bandgap (~1.65 eV), with high absorption in the visible and a weaker but finite absorption in the NIR. In this case, carriers generated close to the WSe$_2$ contacts are directly collected, while those generated deeper within the 2D sheet can migrate laterally and be extracted via the NiO nanowire because the contacts are positioned to bridge both materials, as shown in case (ii) of **Fig. 4a**. Importantly, the difference in work functions between NiO and WSe$_2$ produces an internal electric field at the heterointerface, which aids in exciton dissociation and suppresses recombination losses. Furthermore, the extended coverage area of the 2D WSe$_2$ provides a large active volume for light absorption, especially in the visible regime, thereby generating a high density of excitons that contribute to the device's extraordinary responsivity.

The following assumptions allow a description for the observed dependencies (**Fig 3a-c**). The NiO nanowires and WSe2 monolayer behave like a photoconductor without Schottky-contacts at the gold probes. The generated carriers do not affect the electronic band structure and optical properties of the materials. The generation rate is proportional the electric field intensity $|E|^2$. For the case of photon energy below the bandgap $h\nu < E_g$ there is also an exponential term $\exp\left((h\nu - E_g)/kT\right)$. Under these assumptions, we can evaluate the stationary photocurrent from the balance between the generation and recombination rates. The unknown parameters of the model are subject to fitting the experimental data.

## 3.4. Nanophotonic Enhancement and Broadband Response

The transport analysis above explains which material predominantly contributes carriers in each spectral window; however, it does not account for the magnitude of the measured responsivity, which is unexpectedly large even in regimes where an atomically thin $WSe_2$ sheet would be expected to absorb only weakly. To address this, we explicitly treat the 1D/2D heterostructure as a nanophotonic element. The single NiO nanowire with a diameter of 200 nm is a high-index dielectric body that is essentially subwavelength across the operating band. Such structures are known to support Mie-type resonance modes [37]. A finite-element solutions of Maxwell's equations show that these modes concentrate the optical electric field into a nanoscale mode volume located at, and overlapping with, the $WSe_2$/NiO junction [38, 39]. The hot spot that is a local enhancement in the time-averaged electromagnetic energy density $|\mathbf{E}|^2$ is located at the junction and strongly enhances the effective optical absorption cross-section [40]. We quantify this effect via a "photon localization spectrum," defined by integrating $|\mathbf{E}|^2$ over the interfacial nanojunction volume [**Fig. 4b)**]. The experimentally observed peaks in responsivity coincide with peaks in this photon localization spectrum, indicating that the dominant photocarrier generation pathway is set by a nanophotonic near-field concentration mechanism rather than by uniform bulk absorption in either constituent material.

In this picture, the device can be understood as a nanoscale absorber system boosted with an NiO low-loss optical antenna. It captures incident wave and concentrates optical field through the geometry-defined Mie-type mode, rather than through a plasmonic free-carrier resonance. The atomically thin $WSe_2$ in direct contact with the nanowire acts as the absorber placed at the field hot-spot area, where the local optical intensity (and therefore the photocarrier generation rate) is maximized. This spatial funneling of electromagnetic energy effectively increases the absorption cross-section of the junction and enables high internal quantum efficiency even though the active volume is nanometric in thickness. Because the underlying resonance is dielectric and displacement-current–dominated, rather than plasmonic, the near-field enhancement persists across a broad band. We calculated strong photon localization extending from the near-UV through the visible and into the near-infrared, consistent with the experimentally observed broadband responsivity. We incorporate this localization directly into a coupled optoelectronic model by using the simulated $\int |E|^2 \, dV$ at the junction as the optical generation term that seeds a carrier rate-equation model (capture, detrapping, extraction) [41]. This quantitatively reproduces the measured responsivity spectrum and its incident-power scaling using only two electronic fit parameters, indicating that the dominant gain pathway is photonic in origin (explained in detail in the Supplementary Information file).

To directly correlate this optical resonance with electronic output, the photon localization spectrum was used as the primary input to calculate photocarrier generation rates in the coupled model. The theoretical responsivity and photoconductivity obtained from this model quantitatively reproduce the experimental measurements across all three spectral regions [**Fig. S4**], using only two fitting parameters for recombination and trapping. This agreement provides conclusive evidence that nanophotonic resonance is the dominant operative pathway behind the broadband high responsivity.

## 4. Discussion and conclusions

**A refined description of the spectral response can thus be formulated:**

Visible Response (~2.3 eV): The primary Mie resonance of the NiO nanowire strongly amplifies local fields, resulting in enhanced absorption in the $WSe_2$ monolayer and, to a smaller extent, in the NiO itself. This synergistic effect yields extraordinarily high responsivity in the visible region.

Near-Infrared Response (~1.7 eV): Although the photon energy is slightly below the direct bandgap of WSe$_2$, the hot-spot associated with the resonance compensates for the otherwise weak absorption, enabling efficient exciton generation and extraction, thereby sustaining significant NIR response.

Ultraviolet Response (>3.4 eV): In this regime, responsivity is dominated by intrinsic NiO absorption. While nanophotonic resonances are weaker at such high photon energies, modest field localization still contributes to improved carrier generation efficiency.

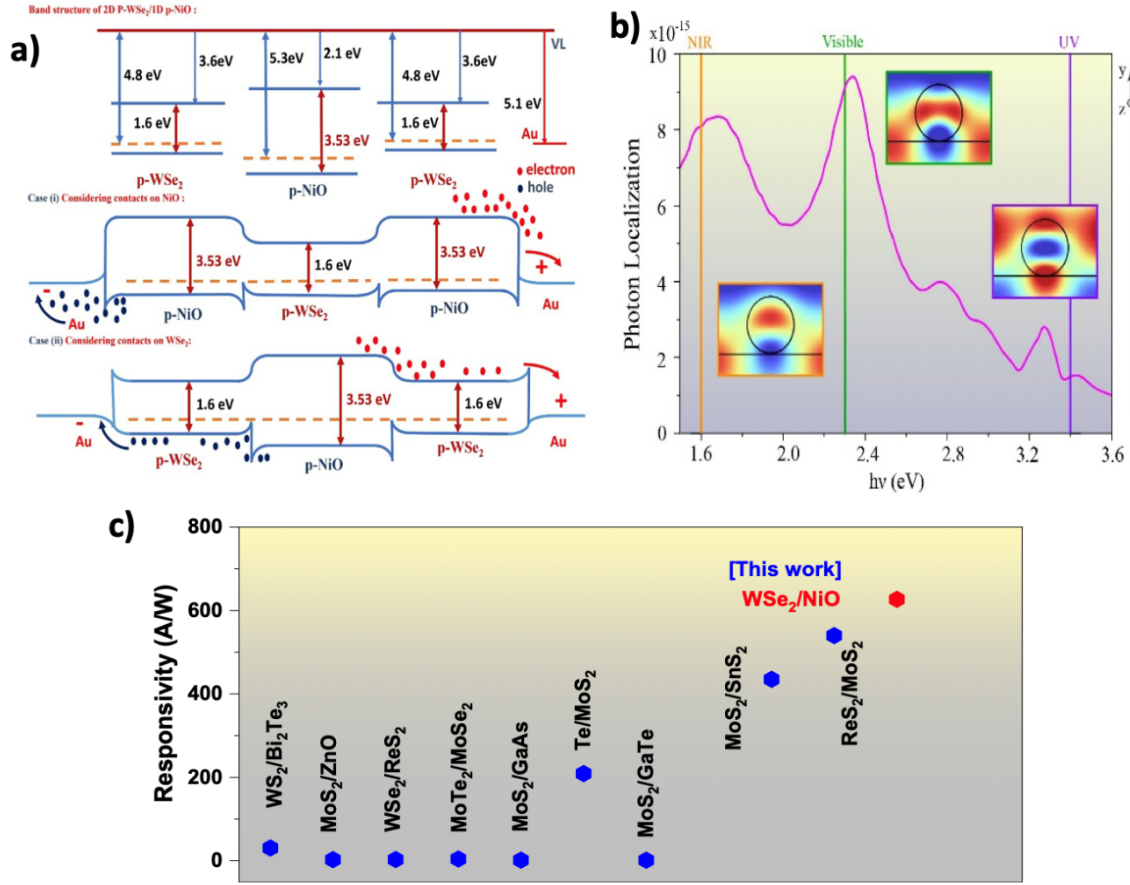

**Fig:4** a) Fig. 4: (a) Band diagrams of the 1D p-NiO/2D p-WSe$_2$ heterojunction. (b) Photon localization spectrum at the heterojunction, with insets showing electric field confinement. (c) Comparison of responsivity achieved in the present work with other reported photodetectors from the literature.

To perform the quantitative comparison with the state of the art, we briefly review other reports on high-responsivity systems. *Yao et al*. fabricated a broadband photodetector with WS$_2$/Bi$_2$Te$_3$ and noted responsivity as 30.7A/W. *Kang et al*. reported on a heterostructure photodetector with MoS$_2$/ZnO by atomic layer deposition with a responsivity of 2.7A/W. *Varghese et al*. demonstrated WSe$_2$/ReS$_2$ based heterostructure photodetector by mechanical exfoliation and claimed the device responsivity of 3 A/W. *Luo et al*. fabricated a heterostructure broadband (360 nm-780 nm) photodetector using MoTe$_2$/MoSe$_2$ by CVD, where the device exhibited responsivity of 4.3A/W. *Wang et al* presented a MoS$_2$/GaAs based heterostructure and observed responsivity of 1.54 A/W. *Park et al*. fabricated a Te/MoS$_2$-based broadband heterostructure by mechanical exfoliation and reported a responsivity of 209A/W. *Mu et al.* showed a MoS2/GaTe-based photodetector that exhibited a responsivity of 1.365 A/W. *Kolli et al.* reported on a mixed dimension broad band photodetector using $MoS_2$/$SnS_2$QDs with a responsivity of 435 A/W. *Polumati et al.* fabricated a mixed dimension broadband photodetector using ReS2/MoS2 QDs and observed responsivity of 540A/W. The proposed mixed dimension heterostructure device with a single 1D p-NiO nanowire and 2D p-WSe2 is capable of delivering high responsivity over a wide range of electromagnetic spectra compared with many reported values mentioned above, and the same is shown in **Fig.4e** below and **Table S1** of SI.

In summary, we have reported on the successful fabrication of a mixed-dimensional broadband photodetector by integrating electrospun 1D p-NiO nanowires with CVD-grown 2D p-$WSe_2$ on Si/$SiO_2$ substrates. The device achieves exceptional responsivities of 627, 227, and 166 A/W under visible, UV, and NIR illumination, respectively, outperforming the systems reported to date. Crucially, our study has provided a definitive elucidation of the underlying physics, establishing that the superior broadband response arises from nanophotonic enhancement rather than a mere superposition of material properties. The subwavelength NiO nanowire functions as a resonant dielectric antenna supporting Mie-type photonic modes that confine the electric field at the 1D/2D nanojunction. This resonant localization has been validated by a coupled optoelectronic model and has been shown to be the dominant pathway for efficient photocarrier generation. By demonstrating the geometry-driven optical enhancement in mixed-dimensional heterostructures, our work has introduced a generalizable design strategy for next-generation ultra-sensitive optoelectronic platforms in imaging, on-chip communication, and environmental sensing.

## Acknowledgements:

C.S.K.R. would like to acknowledge funding from Anusandhan National Research Foundation (ANRF) NPDF (Grant No. PDF/2023/003336). A.A.K., E.E.M., M.V.R. acknowledge support from the Russian Scientific Foundation (Grant No. 25-12-00213).

## Competing interests

The authors declare no competing interests.

## Data Availability Statement

The data that support the findings of the research are available from the corresponding author upon reasonable request.